\documentclass{ifacconf}

\usepackage{graphicx}      
\usepackage{natbib}        
\usepackage{mathrsfs}
\usepackage{mathtools}
\usepackage{amsmath}
\usepackage{amsfonts}
\usepackage{booktabs}
\usepackage{diagbox}
\usepackage{pifont}
\begin{document}
\begin{frontmatter}

\title{Bayesian Model Selection of Lithium-Ion Battery Models via Bayesian Quadrature\thanksref{footnoteinfo}}

\thanks[footnoteinfo]{This work has been accepted to IFAC 2023. Code is publicly available at: https://github.com/Battery-Intelligence-Lab/BayesianModelSelection. The authors acknowledge support by Toyota Motor Corporation, Oxford Clarendon Fund, Oxford Kobe scholarship, the German Aerospace Center (DLR), and the Helmholtz Association through grant no KW-BASF-6,
and contributes to the research performed at CELEST}

\author[First]{Masaki Adachi}$^{, **}$
\author[Fourth]{Yannick Kuhn}$^{, ****}$
\author[Fourth]{Birger Horstmann}$^{, ****}$
\author[Fourth]{Arnulf Latz}$^{, ****}$
\author[First]{Michael A.\ Osborne}
\author[Second]{David A.\ Howey}$^{, \dagger}$

\address[First]{Machine Learning Research Group, University of Oxford, OX2 6ED, UK (e-mail: masaki@robots.ox.ac.uk).}
\address[Second]{Battery Intelligence Lab, University of Oxford, OX1 3PJ, UK}
\address[Fourth]{German Aerospace Center (DLR), Pfaffenwaldring 38-40, 70569 Stuttgart, Germany\\Helmholtz Institute Ulm, Helmholtzstraße 11, 89081 Ulm, Germany}
\address[Fifth]{Universität Ulm, Albert-Einstein-Allee 47, 89081 Ulm, Germany}
\address[sixth]{The Faraday Institution, Harwell Campus, Didcot OX11 0RA, UK}

\begin{abstract}                
A wide variety of battery models are available, and it is not always obvious which model `best' describes a dataset. This paper presents a Bayesian model selection approach using Bayesian quadrature. The model evidence is adopted as the selection metric, choosing the simplest model that describes the data, in the spirit of Occam's razor. However, estimating this requires integral computations over parameter space, which is usually prohibitively expensive. Bayesian quadrature offers sample-efficient integration via model-based inference that minimises the number of battery model evaluations. The posterior distribution of model parameters can also be inferred as a byproduct without further computation. Here, the simplest lithium-ion battery models, equivalent circuit models, were used to analyse the sensitivity of the selection criterion to given different datasets and model configurations. We show that popular model selection criteria, such as root-mean-square error and Bayesian information criterion, can fail to select a parsimonious model in the case of a multimodal posterior. The model evidence can spot the optimal model in such cases, simultaneously providing the variance of the evidence inference itself as an indication of confidence. We also show that Bayesian quadrature can compute the evidence faster than popular Monte Carlo based solvers.
\end{abstract}


\end{frontmatter}

\section{Introduction}
The lithium-ion battery is key to decarbonising power grids and electrifying vehicles. However, its behaviour can be challenging to model, control, and diagnose, and this is a practical hindrance to obtaining the optimal performance. This is compounded by the available data from operational batteries being typically limited to just three measurements: voltage, current, and temperature. Estimating internal states from these time-varying three variables is challenging or even mathematically impossible due to parameter identifiability issues \citep{bizeray2018identifiability}. Degradation further complicates matters since the number of parameters to be identified becomes larger when considering long-term ageing.

There are dozens of plausible models for Li-ion batteries, owing to differing assumptions and levels of approximation. While electrochemists might prefer continuum models, such as the Doyle-Fuller-Newman model \citep{doyle1993modeling}, that give understanding of internal chemical reactions and transport, control engineers prefer simpler approaches such as equivalent circuit models (ECMs) \citep{he2011evaluation}, for fast control and fewer parameters. Other models exist in a spectrum between these (e.g.\ from simple to more complex: ECM $\rightarrow$ EHM \citep{milocco2014generic} $\rightarrow$  SPM \citep{santhanagopalan2006review} $\rightarrow$  SPMe \citep{kemper2013extended} $\rightarrow$  DFN). System identification is the foundation of an estimation and control system, determining predictive accuracy, quick response, and reliability. 

However, the `best' model should be ascertained based on quantifiable performance metrics. Importantly, the optimal model strongly depends on the dataset \textbf{D} and user requirements. A widely accepted approach for defining `good' models is Occam's razor, where the simplest model to reasonably reproduce a given dataset is considered the best. Simplest here relates to the number of parameters to be identified. \cite{rasmussen2000occam} showed that such a metric could be evaluated via Bayesian model evidence, obtained for a model $M$ by integrating out (i.e.\ averaging over) the parameters $\theta$ from the likelihood,
\begin{align}
    p(\textbf{D} | M) = \int p(\textbf{D} | \theta, M) \text{d}p(\theta),
\end{align}
where $p(\theta)$ is the prior distribution and $p(\textbf{D} | \theta, M)$ is the likelihood. The mean evidence $\mathbb{E}[p(\textbf{D} | M)]$ gives the probability of reproducing a given dataset $\textbf{D}$ with a given model $M$, 
the degree of model fit penalised by model complexity. The variance quantifies its uncertainty.

Surprisingly, 
Bayesian model selection of battery models has barely been reported, except for \cite{miyazaki2020bayesian}. Although Bayesian parameter estimation \citep{aitio2020bayesian, escalante2021uncertainty}, and probabilistic modelling works \citep{huang2021towards, liu2020gaussian} exist, most Bayesian approaches in the battery community use Markov chain Monte Carlo (MCMC) \citep{metropolis1953equation, hasting1970monte}, a user-friendly but sample-inefficient approach for inference. Recent work \citep{kuhn2022ep} on parameterisation applied a sample-efficient solver with Bayesian optimisation, none of the above solvers offer evidence computation. This is because estimating the evidence requires prohibitive integral computation, and this is particularly challenging when the likelihood is non-closed-form and/or expensive. A typical practice in such cases is to adopt the Bayesian information criterion (BIC), which is a coarse approximation of the evidence that assumes the posterior is a unimodal Gaussian. Unfortunately battery parameter estimation can produce multimodal or non-Gaussian posterior distributions \citep{aitio2020bayesian, escalante2021uncertainty}, and ignoring this may cause overconfidence---previous work \citep{miyazaki2020bayesian} demonstrates that the identification of the best model using a variant of BIC gradually worsens as the posterior multimodality increases. This paper introduces Bayesian quadrature (BQ) as a novel technique for sample-efficient model evidence \textit{and} parameter posterior estimation, and applies this to battery equivalent circuit models using synthetic data for demonstration purposes.

\section{Battery Model Formulation}
\begin{figure}
\begin{center}
\includegraphics[width=8.4cm]{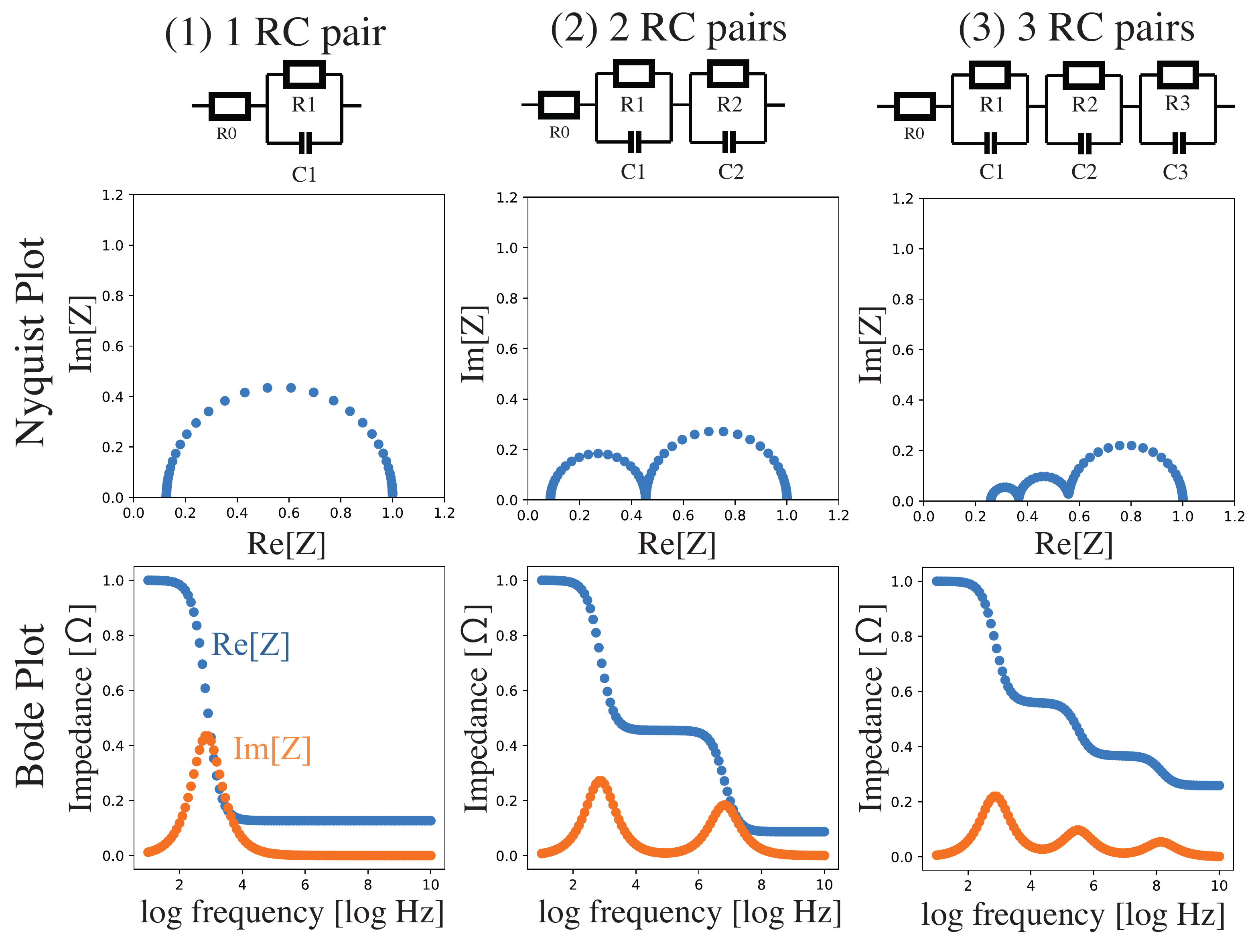}
\caption{Model selection from three RC pair models.}
\label{fig:ECM}
\end{center}
\end{figure}

We selected ECMs for proof-of-concept here since they are relatively simple battery models that nonetheless offer identification challenges. Parameter identifiability for ECMs is often examined in the frequency domain, for example via electrochemical impedance spectroscopy (EIS) data, although time domain data may also be used. Several plausible ECMs are usually compared when fitting EIS data, but the process is subjective and based on the user's electrochemical understanding of the target battery. For simplicity, we chose a simple resistance-capacitance (RC) pair model---this may represent various physical processes, for example kinetics and double layer capacitance, or an approximation of diffusion. Fig.\ \ref{fig:ECM} illustrates the circuit configurations and typical Nyquist plots of three variations of RC circuit models. The number of RC parallel connection components corresponds to the number of the semi-circles in a Nyquist plot. This correspondence is key for identifying the model from spectra. As the semi-circle shape implies, the real and imaginary parts of spectra have a mathematical relationship (Kramers-Kronig relations in Debye relaxation), where one part of spectra can be derived from the other via an equation.

We extend this formulation to make the model better suited for statistical inference using the hyperbolic formulation \citep{calderwood2003physical}; we improve this, permitting non-dimensionalised parameterisation without positivity constraint, as follows. For a general circuit with $N$ total RC pairs plus an additional series resistance $R_0$, where $R_i$ is the resistance of $i$-th RC pair [$\Omega$], $\ln ( \omega \tau_i)$ is the rescaled frequency scale to make the scale independent of the given frequency range of the dataset, rescaled with the breakpoint frequency $\omega_i := 1/\tau_i$ [rad/s], $\tau_i := R_i C_i$ is the time constant of the $i$-th RC pair [s], $C_i$ is the capacitance of the $i$-th RC-pair [F], $f$ is the frequency [Hz] and $\omega := 2\pi f$ is the angular frequency [rad/s], one can define the total resistance $R_\text{total}$, the log of this $r_\text{total}$ (which is positive), the dimensionless resistance of $i$-th RC-pair $r_i$ (constrained between zero and one), the unconstrained dimensionless resistance $r_i^\prime$, the scaling factor $R_\text{im}$, and the weight of $i$-th hyperbolic secant distribution $w_i$, as follows:
\begin{align}
    R_\text{re} &:= R_0 + \sum_{i=1}^N R_i := R_\text{total} := \exp(r_\text{total}),\\
    r_i &:= \frac{R_i}{R_\text{total}} := \exp\left[ -\exp(r_i^\prime) \right],\\
    R_\text{im} &:= \frac{\pi}{2}\sum_{i=1}^N R_i \\
    \quad \lambda_i &:= \frac{R_i}{\sum_{i=1}^N R_i}.
\end{align}
From this, the real and imaginary parts of the impedance ($\text{Re}[Z]$, $\text{Im}[Z]$) , are given by (see Appendix \ref{sec:ecm})
\begin{align}
    \text{Re}[Z] &= R_\text{re} \left[ r_0 + \sum_{i=1}^N \frac{r_i}{2} \left[ 1 - \tanh(\ln \omega \tau_i) \right] \right],\label{eq:ReZ}\\
    \text{Im}[Z] &= \underbrace{R_\text{im}}_\text{scaling factor} \underbrace{\left[ \sum_{i=1}^N \frac{\lambda_i}{\pi} \text{sech}\left(\ln \omega \tau_i\right)\right]}_{\substack{\text{mixture of hyperbolic} \\ \text{secant distributions}}} \label{eq:ImZ}.
\end{align}

The frequency range is also standardised according to the frequencies in the available dataset, 
\begin{align}
    \mu_\omega, \sigma_\omega &:= \mathbb{E}[\ln \omega], \sqrt{\mathbb{V}\text{ar}[\ln \omega]},\\
    \omega^\text{std}, \tau^\text{std}_i &:= \frac{\ln \omega - \mu_\omega}{\sigma_\omega},  -\frac{\ln \tau_i + \mu_\omega}{\sigma_\omega},
\end{align}
where the mean $\mu_\omega$ and standard deviation $\sigma_\omega$ of logarithmic angular frequency\footnote{Where necessary we assume arguments of logarithms are divided by appropriate units, e.g.\ 1 [rad/s], to ensure they are dimensionless.}$, \ln \omega$ can be calculated from the given frequency range of the dataset, and from this we define a standardised frequency scale $\omega^\text{std}$ and standardised time constants $\tau^\text{std}_i$. These may be related to the actual time constants and capacitances (noting that $\tau_i$ is the unstandardised form of the time constant) via 
\begin{align}
    \ln \omega \tau_i &:= \ln \omega - \sigma_\omega \tau^\text{std}_i - \mu_\omega,\\
    C_i &= \frac{\tau_i}{r_i R_\text{total}}.
\end{align}

The parameters to be fitted are unconstrained standardised ones $\theta := \{ r_\text{total}, r^\prime_i, \tau^\text{std}_i\}$. This formulation is similar to the distribution of relaxation times modelling. 
This canonical form provides three benefits: separation of scaling factor, unconstrained prior distribution selection for all parameters, and integral-friendly formulation. Separating the scaling factors can decompose parameter estimation problems into problems of estimating magnitudes ($R_\text{re}$) and ratios ($r_i$), permitting fair comparison over varied magnitudes of resistance. Logarithmically transformed parameters enable non-negativity constraints over resistance, allowing arbitrary prior distributions to be used for Bayesian inference (for instance, $r_i$ is constrained between zero and one, but $r^\prime_i$ is unconstrained). The mixture of hyperbolic secant distributions offers several integral identities to analytically calculate the expectation and variance (see Appendix \ref{sec:ident}). Moreover, this formulation interprets the imaginary part as a probability distribution function, allowing statistical analysis (see section \ref{sec:identifiability}).

\section{Bayesian Inference Formulation}
We wish to select the likeliest model from the above-mentioned three RC pair options. In Bayesian inference, we need to assume a prior distribution $p(\Theta) := \pi(\Theta)$ and a likelihood function $p(\textbf{D}|\Theta, M) : = \ell_\text{true}(\Theta)$. The prior distribution is a probability distribution reflecting one's prior assumptions about possible parameters. For instance, we adopt here a multivariate normal distribution $\pi(\Theta) := \mathcal{N}(\Theta; \mu_\pi, \boldsymbol\Sigma_\pi)$. The mean vector $\mu_\pi$ represents our guess of plausible parameter values and the covariance matrix $\boldsymbol\Sigma_\pi$ reflects our assumption on the uncertainty of each parameter, and correlations between parameters. The likelihood function $\ell_\text{true}(\Theta)$ is a probability distribution to evaluate how the selected parameter set $\Theta$ can reproduce the given dataset $\textbf{D}$. Here we assume a univariate Gaussian with zero mean $\textbf{0}$ and homoskedastic noise, meaning the noise variance $\sigma_\text{noise}$ does not vary over frequency. The squared error evaluates how similar the observed data $y_\text{obs}$ and ECM predicted data $y_\text{ecm}$ are. Now, with the assumed prior $p(\Theta)$ and likelihood function $p(\textbf{D}|\Theta, M)$, Bayes' rule defines the parameter posterior as $p(\Theta, M|\textbf{D})$ and the model evidence $p(\textbf{D}|M)$, all as follows:
\begin{align}
    p(\textbf{D} | \Theta, M) &:= \ell_\text{true}(\Theta) := \prod_j^m \mathcal{N}(\text{err}_j(\theta); \textbf{0}, \sigma^2_\text{noise}),\\
    p(\textbf{D} | M) &:= \mathcal{N} \left( \mathbb{E}_{\pi}[\ell_\text{true}(\Theta)], 
    \mathbb{V}\text{ar}_{\pi}[\ell_\text{true}(\Theta)] \right), \label{eq:evidence_}\\
    p(\Theta | \textbf{D}, M) &= \frac{p(\textbf{D} | \Theta, M) p(\Theta)}{p(\textbf{D} | M)} = \frac{\ell_\text{true}(\Theta)\pi(\Theta)}{\mathbb{E}_{\pi}[\ell_\text{true}(\Theta)]}, \label{eq:posterior}
\end{align}
where
\begin{align}
    \textbf{D} &:= \{\textbf{y}_\text{obs}, \omega^\text{std} \} \in \mathbb{R}^{m \times 2},\\
    \theta &:= \{ r_\text{total}, r^\prime_i, \tau^\text{std}_i\} \in \mathbb{R}^{d-1},\\
    \Theta &:= \{ \theta, \sigma^2_\text{noise} \} \in \mathbb{R}^{d},\\
    y_{\text{ecm}, j}(\theta) &:= \{ y_{\text{re}, j}, y_{\text{im}, j} \} = M(\theta, \omega^\text{std}_j),\\
    \text{err}_j(\theta) &:= \left[y_{\text{obs}, j} - y_{\text{ecm}, j}(\theta) \right]^2,
\end{align}
where subscript `obs' refers to measured data, subscript `ecm' to modelled data, and $M$ is the model (equations (\ref{eq:ReZ})- (\ref{eq:ImZ})). The posterior $p(\Theta|\textbf{D}, M)$ is a conditional probability distribution that reflects our updated estimate of the parameter space based on the observed data $\textbf{D}$. We use dimensionless and unconstrained $r^\prime_i$ and $\tau^\text{std}_i$ as inputs of the model for arbitrary prior selection and fair comparison of models. The number of parameters to be estimated is $d = 2 + 2N$, as the scaling factor $r_\text{total}$ and experimental noise variance $\sigma^2_\text{noise}$ are shared over all models.

\section{Bayesian quadrature modelling}\label{sec:BayesInf}
We wish to estimate both the parameter posterior distribution $p(\Theta|\textbf{D}, M)$ and the evidence $p(\textbf{D}|M)$. We also wish to minimise the number of times that the likelihood $\ell_\text{true}(\Theta)$ must be queried, as this could be a computationally demanding operation in a more complex model. This problem requires a sample-efficient Bayesian inference solver. Bayesian quadrature (BQ) offers sample efficiency and solves for the posterior and the evidence in one go. This is a surrogate-model-based numerical integration approach, solving the integral as an inference problem by modelling the likelihood function $\ell_\text{true}(\Theta)$ with a Gaussian process (GP). Define $\ell(\Theta)$ as the surrogate likelihood function modelled by a GP. The key result is that BQ can recast the problem of Bayesian inference into one of function approximation. The more accurately $\ell(\Theta)$ can predict $\ell_\text{true}(\Theta)$, the more accurately the posterior and evidence can be estimated via replacing $\ell_\text{true}(\Theta)$ with $\ell(\Theta)$ in Eqs.\ (\ref{eq:evidence_}) - (\ref{eq:posterior}). To achieve this, \cite{adachi2022fast} proposed \textit{BASQ}, a discrete approximation of the kernel integral using a kernel recombination method \citep{hayakawa2022positively}, yielding the following evidence computations:
\begin{align}
    \text{LEM} &:= \ln \mathbb{E}_{\pi}[\ell(\Theta)] \approx \ln \sum_k^L W_k \mu_f(X_k) + \beta, \label{eq:lem}\\
    \text{LEV} &:= \ln \mathbb{V}\text{ar}_{\pi}[\ell(\Theta)] \approx \ln \sum_{k,l}^L W_k W_l \sigma_f(X_k, X_l) + 2\beta,\label{eq:lev}
\end{align}
where LEM and LEV refer to log evidence mean and log evidence variance, $\mu_f$ and $\sigma_f$ are the predictive mean and covariance of the likelihood surrogate model $\ell(\Theta)$, $\beta$ is the scaling constant, $W_k, W_l$ and $X_k, X_l$ are the positive weights and point configurations discretised by the kernel recombination. Recall that LEM gives the degree of model fit and the LEV quantifies the uncertainty of the fit. However, the prior work on this \citep{adachi2022fast} assumed a narrower dynamic range of likelihood, whereas the battery model typically produces $10^{700}$ likelihood values. This is way beyond a typical numerical overflow limit. Thus, we improve here the prior work by adopting a four-layered warped GP method to accommodate the wide dynamic range of likelihood. (See Appendix \ref{sec:BQ})

\section{Identifiability}\label{sec:identifiability}
To evaluate the model evidence as a model selection criterion, we compare results against three classical metrics related to identifiability: number of data points $m$, signal-to-noise ratio (SNR), and Jensen-Shannon divergence (JS). Owing to the integral-friendly model formulation, most parts of these can be calculated analytically. The number of data points is controllable here because data are synthetically generated and  equispaced over log angular frequency space. Both SNR and JS are calculated using the imaginary part of the impedance. As the canonical form can be regarded as a mixture of hyperbolic secant distributions, such statistical analysis can be applied. While SNR evaluates the identifiability along the impedance magnitude axis, JS does so along the frequency axis.

\subsection{Signal-to-noise ratio}
The SNR is the log fraction of the impedance variance over the noise variance, representing how much the signal is more distinct than the noise, defined as:
\begin{align}
    \text{SNR} &:= \ln\frac{\mathbb{V}\text{ar}_{P(\ln \omega)}[\text{Im}[Z]]}{\sigma^2_\text{noise}}.
\end{align}
Larger SNR means a more distinct and identifiable signal. The canonical form of the model provides an analytical form for the SNR (see derivation in Appendix \ref{sec:snr}).

\subsection{Jensen-Shannon divergence}
The JS divergence is a distance metric quantifying how one probability distribution $P_i(x)$ is similar to a second reference probability distribution $P_j(x)$, defined as:
\begin{align}
    \begin{split}
    \text{JS} &:= \frac{1}{2}\int \ln \left( \frac{P_i(x)}{M_{ij}(x)}
    \right) \text{d} P_i(x) \\
    &+ \frac{1}{2}\int \ln \left( \frac{P_j(x)}{M_{ij}(x)}
    \right) \text{d} P_j(x),\\
    \end{split}
\end{align}
where
\begin{align}
    M_{ij}(x) &:= \frac{1}{2}\Big(
    P_i(x)+P_j(x) \Big)
\end{align}
As the JS is defined for pairwise comparisons, the number of criteria required increases combinatorially per the number of RC pairs. For simplicity, we only consider the case of two RC pairs, which produces only one JS divergence. This represents how much the selected two peaks in the imaginary parts overlap. A smaller JS divergence means a more distinguishable and identifiable signal. While SNR is determined by the noise variance $\sigma^2_\text{noise}$ and scaling factor $r_\text{total}$, JS is dominated by the time constant difference $\Delta \tau_{ij}$. Again, the canonical form helps solve the integration. Note that this is formulated as noise-free. The integration calculation procedure can be seen in Appendix \ref{sec:JS}. The extended JS to include noise $\sigma^2_\text{noise}$ is also guided, but the results shown in this paper are consistently used with noise-free formulation for simplicity. 

\section{Numerical Results}
\subsection{Selection criteria comparison}\label{sec:easyhard}
We now demonstrate our modified version of BASQ over several cases. We compare the model evidence metric (LEM and LEV\footnote{LEV values in the tables are standardised via subtracting $2\beta$ from eq. (\ref{eq:lev}) for a fair comparison between models.}
, eqs. (\ref{eq:lem}) - (\ref{eq:lev})) with root-mean-square error (RMSE), Bayesian information criterion (BIC), and expected log predictive density (ELPD), based on the maximum a posteriori (MAP) parameter estimates, defined as:
\begin{align}
    \Theta_\text{MAP} &:= \text{argmax} \, \ell_\text{true}(\Theta),\\
    \text{RMSE} &:= \sqrt{\frac{1}{m}\sum_j^m \text{err}_j(\theta_\text{MAP})},\\
    \text{BIC} &:= d \ln m - 2 \ln \ell_\text{true}(\Theta_\text{MAP}),\\
    \text{ELPD} &:= \sum_j^m \ln \int \ell_\text{true}(\Theta) \text{d} p(\Theta| \textbf{D}, M).
\end{align}
The RMSE is a noise-free formulation that does not consider parameter uncertainty. BIC is an asymptotic approximation of evidence, so it cannot evaluate multimodal likelihoods. ELPD is a similar formulation to the log mean evidence, but the probability measure is changed from prior to posterior. The motivation behind ELPD is to estimate the alternative evidence from MCMC samples, as it cannot estimate evidence when solving Bayesian inference. However, it relies on Monte Carlo (MC) integration, which requires a significant amount of posterior samples, meaning that a plethora of model evaluations $\ell_\text{true}(\Theta)$ will run. All these alternative criteria were calculated from the BQ estimated posteriors by post-processing. Moreover, none of these criteria quantify their own uncertainty except BQ.
\begin{table}
  \caption{Easy case}
  \label{tab:easy}
  \centering
  \renewcommand{\arraystretch}{1.0}
    \begin{tabular}{lcccc}\toprule
     &\makebox[3em]{1 RC pair}&\makebox[3em]{2 RC pairs}&\makebox[3em]{3 RC Pairs}&\makebox[3em]{4 RC pairs}\\
    true model &  & \ding{52} & & \\\midrule
    LEM & -2809233 & \textbf{703.6569} & 289.2976 & 225.1602\\
    LEV & -33.52068 & -27.31169 & -31.91129 & -31.38766 \\\midrule
    RMSE & 1.147527 & \textbf{0.006677} & 0.031770 & 0.062151 \\
    BIC & 5766999 & \textbf{-1405.553} & -572.7390 & -432.4597 \\
    ELPD & -2883641 & \textbf{713.7332} & 293.2492 & 213.7417 \\\bottomrule
    \end{tabular}
\end{table}

\begin{table}
  \caption{Hard case}
  \label{tab:hard}
  \centering
  \renewcommand{\arraystretch}{1.0}
    \begin{tabular}{lcccc}\toprule
     &\makebox[3em]{1 RC pair}&\makebox[3em]{2 RC pairs}&\makebox[3em]{3 RC Pairs}&\makebox[3em]{4 RC pairs}\\
    true model & & & \ding{52} & \\\midrule
    LEM & -150.3634 & -151.8002 & \textbf{-147.4257} & -151.9208\\
    LEV & -15.74094 & -19.07997 & -19.45956 & -26.57386 \\\midrule
    RMSE & \textbf{0.492191} & 0.492643 & 0.492269 & 0.492260\\
    BIC & \textbf{302.8601} & 313.8921 & 321.0089 & 310.8289 \\
    ELPD & \textbf{-145.7505} & -148.8758 & -148.4754 & -146.1485\\\bottomrule
    \end{tabular}
\end{table}

We demonstrate the behaviours of the selection criteria on two different datasets---an easy case ($\Delta \tau_{ij}=9.1$, $\ln \sigma^2_\text{noise}=-9.97$) with results in Table \ref{tab:easy}, and a hard case ($\Delta \tau_{ij}=0.36$, $\ln \sigma^2_\text{noise}=-1.6$) detailed in Table \ref{tab:hard}. The easy case is clean data generated with 2 well-separated semi-circles, and the hard case is noisy data generated with an additional third semi-circle with more overlap. The "better" column shows which upward or downward direction is better for each criterion. As expected, separated peaks (large $\Delta \tau_{ij}$) and lower noise $\sigma^2_\text{noise}$ boost identifiability.
While all criteria selected the true model in the easy case, only the evidence can select the true model in the hard case.

The other metrics were unsuccessful in the hard case because of a multimodal posterior in the one RC pair model. As three RC pairs were used to generate the dataset, the posterior distribution of one RC pair parameter inevitably becomes multimodal, such as the peak intensity ($\lambda_i$). While the evidence correctly incorporates the multimodal distribution shape, RMSE and BIC consider only the largest peak. The BIC estimates the whole distribution from the local curvature at the maximum, which becomes erroneously overconfident in the multimodal case \citep{murphy2012machine}. 
ELPD's failure could be due to its rough integral approximation. As the convergence rate of MC integration is $\mathcal{O}(1/\sqrt{n})$, the posterior samples ($n$ = 1,000) is too few. This means more model evaluations $\ell_\text{true}(\Theta)$ are required, which would not scale to slower simulation models.

In contrast, the evidence can be estimated simultaneously during training. Moreover, the variance of the evidence successfully points out the lower confidence in the one RC pair model in the hard case, suggesting multimodality. This uncertainty over the selection criterion could avoid overconfidence toward a simpler model. Moreover, the evidence variance in the hard case is generally higher than in the easy case. This also tells us that the hard case dataset is almost unidentifiable, suggesting we should not trust these comparisons. For instance, the evidence mean and ELPD for one RC pair in the easy case are much lower than in the hard case. However, the integral variance is the opposite. Thus, only this metric quantifies its own uncertainty, suggesting the dataset or model is less informative. A similar notion can be found in Jeffreys' scale for the Bayes factor \citep{jeffreys1998theory}, which claims the evidence is not strong when the difference between the log evidence of two models is lower than 10. This explains that the hard case is unreliable, as the difference in the log evidence shows insufficient plausibility. Contrary to Jeffreys' scale, log evidence variance is self-contained and does not require the comparison of models. Instead, it can independently spot the unreliability of the estimation.

In such an uncertain case, 
a typical practice is Bayesian model averaging. Rather than selecting one definite model, we sample from a \textit{mixture of models} with probability proportional to their mean evidence. Averaging can boost predictive accuracy and reduce the uncertainty over predictions, where only evidence offers this method.
As such, while the easy cases do not require advanced methods, the evidence with self-check on reliability can assist in deciphering minor differences in hardly identifiable problems. 
\begin{table}
  \caption{Linear correlation matrix}
  \label{tab:corr}
  \centering
  \renewcommand{\arraystretch}{1.0}
    \begin{tabular}{lccccc}\toprule
    factors &\makebox[3em]{$m$}&\makebox[3em]{JS}&\makebox[3em]{SNR}&\makebox[3em]{LEM}&\makebox[3em]{LEV} \\\midrule
    $m$ & - & -0.0268 & 0.0058 & 0.4096 & -0.1806 \\
    JS  & -0.0268 & - & -0.0993 & -0.0297 & 0.2243 \\
    SNR & 0.0058 & -0.0993 & - & 0.7299 & -0.4882 \\
    LEM & 0.4096 & -0.0297 & 0.7299 & - & -0.2867 \\
    LEV & -0.1806 & 0.2243 & -0.4882 & -0.2867 & - \\\bottomrule
    \end{tabular}
\end{table}
\begin{table}
  \caption{Functional ANOVA results}
  \label{tab:anova}
  \centering
  \renewcommand{\arraystretch}{1.0}
    \begin{tabular}{lccc}\toprule
    factors &\makebox[3em]{LEM}&\makebox[3em]{LEV}&\makebox[3em]{residual}\\\midrule
    ($m$) & 0.0012 & 0.0083 & 0.0012\\
    (JS) & \textbf{0.3258} & \textbf{0.3335} & \textbf{0.3397} \\
    (SNR) & 0.2778 & 0.2002 & 0.0825\\
    ($m$, JS) & 0.0028 & 0.0088 & 0.0017\\
    ($m$, SNR) & 0.0599 & 0.1260 & 0.2427\\
    (JS, SNR) & 0.2702 & 0.1969 & 0.0879\\
    ($m$, JS, SNR) & 0.0623 & 0.1264 & 0.2443\\\bottomrule
    \end{tabular}
\end{table}

\subsection{Sensitivity analysis}
A sensitivity analysis of the evidence metric was performed. We generated 1,024 datasets using two-RC-pair models while varying the following five parameters; the number of data points $m$, the scaling factor $r_\text{total}$, the first resistance $r^\prime_1$, the first time constant $\tau^\text{std}_1$, and noise variance $\sigma^2_\text{noise}$. 
We calculated the SNR, JS, and the number of data points $m$ for each dataset. In the first step of the analysis, we compared the linear correlations between the evidence estimates. Table \ref{tab:corr} shows Pearson's correlation coefficients. This result aligns with our intuition---for instance, larger data size $m$ and SNR can boost the evidence LEM and confidence (inverse of LEV). However, while the large correlation of evidence with SNR is instinctive, the small correlation with JS is counterintuitive.

Thus, we further investigated the variance analysis via functional ANOVA \citep{hutter2014efficient}, which models a partition of a functional response according to the main effects and interactions of input parameters. This method can attribute each parameter sensitivity in a non-linear manner. Table \ref{tab:anova} illustrates that the most significant influence over the mean and variance of the evidence is the JS, contrary to the linear correlation results. This can be interpreted as meaning that a smaller JS divergence (more overlapped peaks) destabilises the evidence estimation, resulting in a more considerable variance. This viewpoint is supported by the relatively large negative correlation coefficient between JS and LEV.

Further insights can be obtained via residual analysis. The residual is defined as follows:
\begin{align}
    Z_\text{pred} &:= \text{slope} \times \text{BIC} + \text{intercept},\\
    \text{residual} &:= \left(
    Z_\text{pred} - \log \mathbb{E}_\pi[\mu_e(\Theta)]
    \right)^2.
\end{align}
As the BIC is an approximation of the LEM, the BIC and LEV have a linear relationship. While a linear regression model with BIC can predict log evidence mean reasonably, it fails to predict in hard cases, as shown in the section \ref{sec:easyhard}. Residual refers to the squared error between the BIC and log evidence mean. Table \ref{tab:anova} shows that the residual is mainly caused by the JS divergence and less influenced by SNR or the number of data points $m$. This also suggests that the BIC cannot distinguish between the models with overlapped peaks, namely, a multimodal posterior.

\subsection{Computation efficiency}
\begin{figure}
\begin{center}
\includegraphics[width=8.4cm]{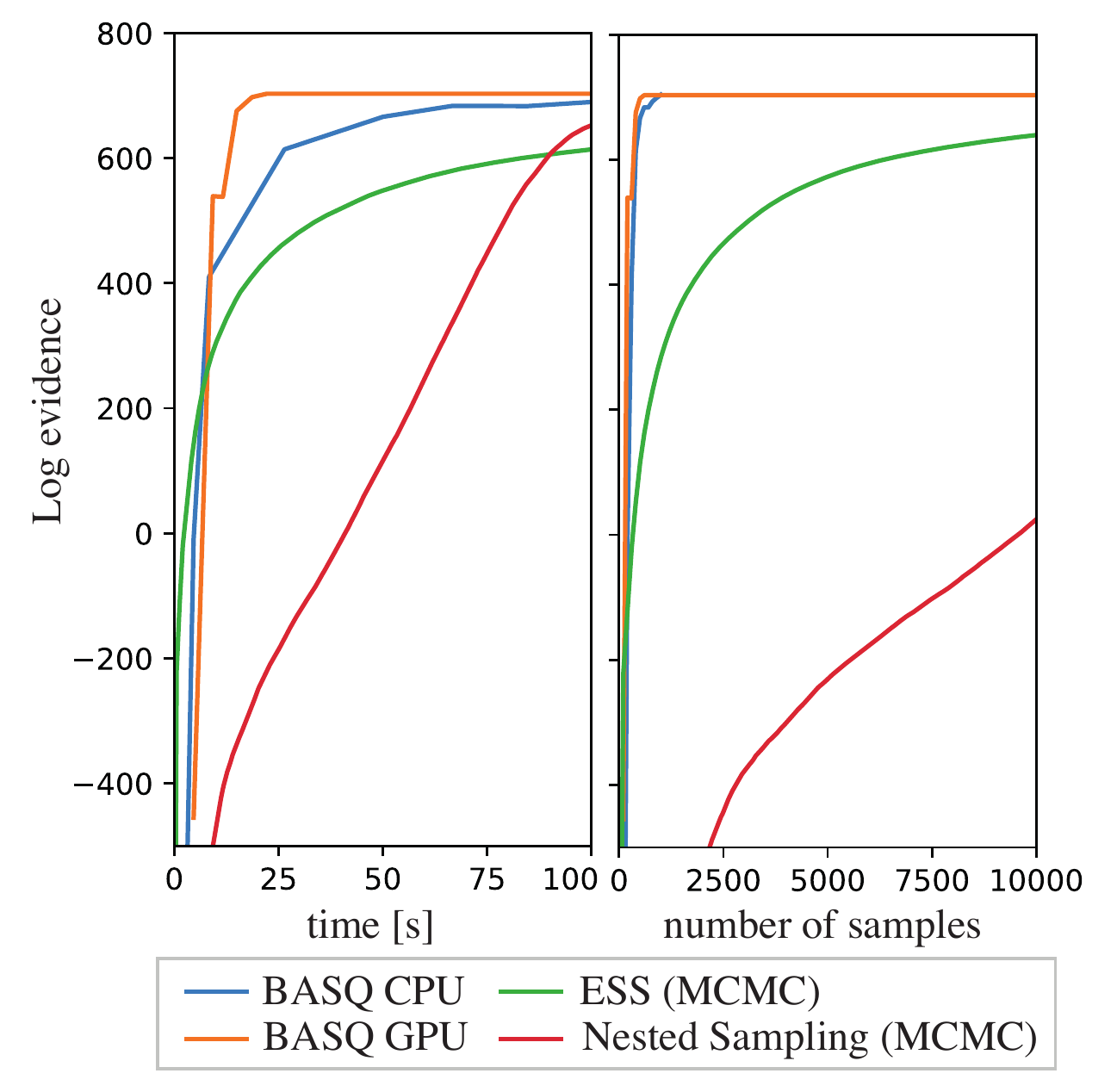}
\caption{The learning curve of log evidence over the computation time and the number of samples.}
\label{fig:comp}
\end{center}
\end{figure}
Lastly, we compared the computation efficiency of our modified version of BASQ with the existing MCMC solvers elliptical slice sampling (ESS) \citep{murray2010elliptical} and dynamic nested sampling \citep{speagle2020dynesty}. Note that amongst MCMC samplers, only nested sampling can estimate the evidence. For ESS, we approximated the evidence using ELPD via posterior samples. Therefore, the estimation with ESS should converge to a larger value than the actual evidence. The BASQ computation was performed using both CPU and GPU.\footnote{
Both MCMC samplers and BASQ in CPU were computed with a MacBook Pro 2019, 2.4 GHz 8-Core Intel Core i9, 64 GB 2667 MHz DDR4. BASQ in GPU was performed on Google Colaboratory.
}

Fig.\ \ref{fig:comp} compares the learning curve of the above four samplers versus computation time, using the easy case dataset shown in Table \ref{tab:easy}. While BASQ in a GPU converges at 18 seconds, BASQ in a CPU converges at 131 seconds. Both ESS and nested sampling do not converge in this time. Fig.\ \ref{fig:comp} contrasts the sample efficiency of the samplers. As BASQ is a parallel sampler, we generate 100 samples per iteration. The sampling efficiency of BASQ does not change over computation modes and is the best of the selected solvers. This is expected---while the convergence rate of BASQ is $\mathcal{O}(\exp(-c n^{1/d})) $ in the Gaussian case \citep{adachi2022fast}, that of MCMC is $\mathcal{O}(1/\sqrt{n})$. Furthermore, even this result does not fully represent BASQ's potential. While ECMs return model predictions in a millisecond order, more complex models (e.g. DFN model) take seconds to query. Therefore, BASQ for such complex models will be even more beneficial. Recent work shows even faster convergence than BASQ \citep{adachi2023sober}.


\bibliography{ifacconf}             

\appendix

\section{Derivation of Canonical form}\label{sec:ecm}
The impedance of RC pair ECM is typically expressed as:
\begin{align*}
    Z &= R_0 + \sum_{i=1}^N \frac{R_i}{1+j\omega C_i R_i},\\
    &= R_0 + \sum_{i=1}^N\frac{R_i}{1 + (\omega C_i R_i)^2} - j \sum_{i=1}^N\frac{\omega C_i R_i^2}{1 + (\omega C_i R_i)^2}.
\end{align*}
This can be written as:
\begin{align*}
    \text{Re}[Z] &= R_0 + \sum_{i=1}^N\frac{R_i}{1 + (\omega C_i R_i)^2},\\
    \text{Im}[Z] &= \sum_{i=1}^N R_i \frac{\omega C_i R_i}{1 + (\omega C_i R_i)^2}.
\end{align*}
For the real part, we can rewrite as:
\begin{align*}
    \text{Re}[Z] &= R_0 + \sum_{i=1}^N\frac{R_i}{1 + (\omega \tau_i)^2},\\
    &= R_0 + \sum_{i=1}^N
    \frac{R_i}{2}
    \frac{2}{1 + (\omega \tau_i)^2},\\
    &= R_0 + \sum_{i=1}^N \frac{R_i}{2}
    \left[
    1 - \frac{(\omega \tau_i)^2 - 1}{(\omega \tau_i)^2 + 1}
    \right],\\
    &= R_0 + \sum_{i=1}^N \frac{R_i}{2}
    \left[
    1 - \tanh(\ln \omega \tau_i)
    \right],\\
    &= R_\text{total}\left[ r_0
    + \sum_{i=1}^N \frac{r_i}{2}
    \left[
    1 - \tanh(\ln \omega \tau_i)
    \right]
    \right],\\
    &= R_\text{re}\left[ r_0
    + \sum_{i=1}^N \frac{r_i}{2}
    \left[
    1 - \tanh(\ln \omega \tau_i)
    \right]
    \right].
\end{align*}
Similarly, the imaginary part can be rewritten as:
\begin{align*}
    \text{Im}[Z] &= \sum_{i=1}^N R_i \frac{\omega \tau_i}{1 + (\omega \tau_i)^2},\\
    &= \sum_{i=1}^N \frac{R_i}{2} \frac{2 \omega \tau_i}{1 + (\omega \tau_i)^2},\\
    &= \sum_{i=1}^N \frac{R_i}{2} \text{sech}(\ln \omega \tau_i),\\
    &= \frac{\pi \left\{ \sum_{i=1}^N R_i \right\}}{2} \sum_{i=1}^N \frac{R_i}{\pi \left\{ \sum_{i=1}^N R_i \right\}} \text{sech}(\ln \omega \tau_i),\\
    &= R_\text{im} \sum_{i=1}^N \frac{\lambda_i}{\pi} \text{sech}(\ln \omega \tau_i),
\end{align*}
where $\lambda_i$ is introduced to be $\sum_{i=1}^N \lambda_i = 1$, $1/\pi$ is introduced to be standardised as $\int_{-\infty}^\infty \frac{1}{\pi} \text{sech}(\ln \omega \tau_i) \text{d} \ln \omega \tau_i = 1$.

\section{Bayesian Quadrature Training Procedure}\label{sec:BQ}
\subsection{Four-layered BASQ formulation}
\begin{table*}
  \caption{Four-layered GPs and warped functions at each layer}
  \label{tab:warpedGPs}
  \centering
  \renewcommand{\arraystretch}{1.7}
    \begin{tabular}{lcccc}\toprule
    Layers &\makebox[3em]{$e$ space}&\makebox[3em]{$f$ space}&\makebox[3em]{$g$ space}
    &\makebox[3em]{$h$ space}\\
    Correspondence & likelihood & normalised likelihood & square-root norm.\ likelihood & sqrt.\ norm.\ log likelihood\\\midrule
    Warp & scaling & square-root & log & base GP\\
    Forward & $e$ & $e/\exp \beta$ & $\sqrt{2(f - \alpha)}$ & $\log(g + 1)$\\
    Backward & $f \exp \beta$ & $\alpha + \frac{1}{2}g^2 $ & $\exp(h) - 1$ & $h$ \\\hline

    GP & $e \sim \mathcal{GP}(\mu_e, \sigma_e)$
    & $f \sim \mathcal{GP}(\mu_f, \sigma_f)$
    & $g \sim \mathcal{GP}(\mu_g, \sigma_g)$
    & $h \sim \mathcal{GP}(\mu_h, \sigma_h)$ \\

    Mean & $\mu_f(x) \exp \beta$
    & $\alpha + \frac{1}{2} \left[ \mu_g(x)^2 + \sigma_g(x,x) \right]$
    & $\exp \left[ \mu_h(x) + \frac{1}{2}\sigma_h(x,x) \right]$
    & $\mu_h(x)$ \\

    Covariance & $\sigma_f(x,y) \exp(2\beta)$
    & $\frac{1}{2} \sigma_g(x,y)^2 + \mu_g(x)\sigma_g(x,y)\mu_g(y)$
    & $\mu_g(x)\mu_g(y)[\exp\{\sigma_h(x,y) - 1 \}]$ 
    & $\sigma_h(x, y)$ \\\bottomrule
    \end{tabular}
\end{table*}

The likelihood surrogate model $\ell(\Theta)$ is defined as:
\begin{align}
    \ell(\Theta) &\sim \mathcal{N}(\ell; \mu_\ell(\Theta), \sigma_\ell(\Theta)),\\
    \mu_\ell(\Theta) &= K(\Theta, \boldsymbol\Theta)K(\boldsymbol\Theta, \boldsymbol\Theta)^{-1} \ell_\text{true}(\boldsymbol\Theta) \label{eq:pred_mean},\\
    \sigma_\ell(\Theta, \Theta^\prime) &= K(\Theta, \Theta^\prime) - K(\Theta, \boldsymbol\Theta) K(\boldsymbol\Theta, \boldsymbol\Theta)^{-1} K(\boldsymbol\Theta, \Theta), \label{eq:pred_var}
\end{align}
where $\ell(\Theta)$ is the surrogate likelihood function modelled by GP, $\boldsymbol\Theta$ is the `observed parameter sets', and $K$ is the kernel.

GP is a non-parametric probabilistic model, typically applied to regression tasks in machine learning. GP can flexibly increase the model complexity in accordance with the number of data, thwarting under/over-confidence. GP model shape is determined by the data points $\boldsymbol\Theta$ and the kernel $K(\Theta, \Theta^\prime)$. The kernel maps the correlation between data points into a covariance matrix. Gaussianity of GP provides analytical predictive distribution $\ell(\Theta)$, with predictive mean $\mu_\ell(\Theta)$ and covariance $\sigma_\ell(\Theta, \Theta^\prime)$, as shown in Eqs (\ref{eq:pred_mean}) - (\ref{eq:pred_var}). While the predictive mean $\mu_\ell(\Theta)$ predicts the likelihood $\ell_\text{true}(\Theta)$, predictive covariance $\sigma_\ell(\Theta, \Theta^\prime)$ predicts the uncertainty of the prediction at given $\Theta$. That is, training GP means minimising the predictive covariance over all possible parameters $\pi(\Theta)$, namely, minimising $\iint_\chi \sigma_\ell(\Theta, \Theta^\prime) \text{d}\pi(\Theta) \text{d}\pi(\Theta^\prime)$. Such training can be done via querying more observations from the true likelihood $\textbf{D}_{\boldsymbol\Theta} = \{\boldsymbol\Theta, \ell_\text{true}(\boldsymbol\Theta)\}$. Hence, the most straightforward training is to sample from the prior $\pi(\Theta)$ until the integral variance becomes smaller than a convergence threshold. However, the prior often barely overlaps over the likelihood, resulting in observing unhelpful tiny likelihood values over most samples.

To overcome this problem, we consider sample-efficient training that fully exploits the information from GP. \cite{osborne2012active} showed that active learning sampling could efficiently reduce the number of samples. The active learning scheme guides the next query point to minimise the integral variance, exploiting the GP surrogate model information. A function called acquisition function formulated by predictive mean $\mu_\ell(\Theta)$ and covariance $\sigma_\ell(\Theta, \Theta^\prime)$ can evaluate where to sample, and optimising it can locate where to sample next. Still, the overhead of the next query guidance is not negligible, and it is an inevitably sequential procedure. \cite{adachi2022fast} proposed batch Bayesian quadrature, termed Bayesian Alternately Subsampled Quadrature (BASQ), permitting a lightweight active learning scheme and parallelisation of querying. They adopted the discretised sampling method \citep{hayakawa2022positively} for probability measure rather than an acquisition function. This allows us to query the true function in parallel. As the modern computational environment exploits an efficient parallel computation via a graphical processing unit or a computer cluster in the cloud, such computing power can accelerate inference computation. They demonstrated that BASQ could accelerate Bayesian inference over various synthetic and real-world datasets, including SPMe model inference.

The evidence can be calculated via kernel recombination. Kernel recombination is a discrete approximation of continuous kernel integral into weighted summation so as to minimise the integral variance, as such:
\begin{align*}
    \int_Q \varphi(x) \text{d} q(x) &\approx \sum_p^P w_p \varphi(X_p),\\
    X_p \in \textbf{X}&, w_p \in \textbf{W}, 
\end{align*}
$\textbf{X}$ is the discretised samples over the probability measure, $\textbf{W}$ is the positive weights to approximate integration. When we recall our training objective is to minimise the predictive covariance over the probability measure $\pi(\Theta)$, this can be formulated as kernel recombination. Hence, we pass the predictive covariance $\sigma_\ell(\Theta, \Theta^\prime)$ as kernel to the kernel recombination algorithm \citep{hayakawa2022positively}, which yields the following approximation:
\begin{align*}
    \textbf{X}, \textbf{W} &= \text{recombination}[\sigma_\ell(\Theta, \Theta^\prime), \pi(\Theta)],\\
    \mathbb{E}_{\pi}[\ell(\Theta)] &= \int_\chi \mu_\ell(\Theta) \text{d}\pi(\Theta),\\
    &\approx \sum_k^L W_k \mu_\ell(X_k),
\end{align*}
\begin{align*}
    \mathbb{V}\text{ar}_{\pi}[\ell(\Theta)] &= \iint_\chi \sigma_\ell(\Theta, \Theta^\prime) \text{d}\pi(\Theta) \text{d}\pi(\Theta^\prime),\\
    &\approx \sum_{k,l}^L W_k W_l \sigma_\ell(X_k, X_l),
\end{align*}
where $X_k, X_l \in \textbf{X}$, $W_k, W_l \in \textbf{W}$.
However, they adopted square-root warping for fast computation, which assumed a narrow dynamic range in likelihood. Battery models' likelihood turns out to be very sharp, as the number of data points over the frequency range is typically over a hundred. 

Therefore, we adopted four-layered GPs to accommodate the dynamic range, permitting solving Bayesian inference even in this wide dynamic range case. Functions at each layer are summarised in Table \ref{tab:warpedGPs}, where $\textbf{Y}_\text{log}$ is the observed log-likelihood values, $\alpha = \min[\exp(\textbf{Y}_\text{log} - \beta)]$, $\beta = \max[\textbf{Y}_\text{log}]$. $e$ space corresponds to the original likelihood space. Square-root warping and log-warping layers are approximated via the moment-matching method \citep{gunter2014sampling, chai2019improving}. To accommodate the wide dynamic range, log transformation is widely applied in the BQ community. However, log-warped GP inevitably results in sampling from log space, leading to ineffective exploration. As meaningful samples from a very sharp likelihood are localised in only the vicinity of the maximum values, log space exploration is too blunt to explore the original space. The combination of square-root warping and log-warping can overcome this issue using the following relationship:
\begin{align*}
    f &= \alpha + \frac{1}{2}g^2 \approx \alpha + \frac{1}{2} \exp(h) \exp(h),\\
    \mathbb{E}_{\pi}[\mu_f(\Theta)] &=\alpha + \frac{1}{2} \int_\Xi \mu_g(\Theta) \text{d} \pi^\prime(\Theta),\\
    \pi^\prime(\Theta) &:= \mu_g(\Theta) \pi(\Theta).
\end{align*}
As such, this doubly warping structure enables us to copy exponentiated function information to both likelihood and prior. Thus, this double structure can sample from sharp exponentiated distribution $\pi^\prime(\Theta)$ as well as keep the surrogate model exponentiated $\mu_g(\Theta)$.

The last layer, $e$, exists to avoid overflow in computation by scaling the whole dynamic range via maximum value. This warping layer can be avoided as such:
\begin{align*}
    \log \mathbb{E}_{\pi}[\mu_e(\Theta)] &= \log \mathbb{E}_{\pi}[\mu_f(\Theta)] + \beta,\\
    &\approx \log \sum_k^L W_k \mu_f(X_k) + \beta,
\end{align*}
\begin{align*}
    \log \mathbb{V}\text{ar}_{\pi}[\sigma_e(\Theta)] &= \log \mathbb{V}\text{ar}_{\pi}[\sigma_f(\Theta)] + 2\beta,\\
    &\approx \log \sum_{k,l}^L W_k W_l \sigma_f(X_k, X_l) + 2\beta,\\
    p(\Theta | \textbf{D}, M) &= \frac{\mu_e(\Theta)\pi(\Theta)}{\mathbb{E}_{\pi}[\mu_e(\Theta)]} =  \frac{\mu_f(\Theta)\pi(\Theta)}{\mathbb{E}_{\pi}[\mu_f(\Theta)]}.
\end{align*}

\subsection{Training procedures}
Training consists of four processes:
\begin{enumerate}
    \item Subsampling from the exponentiated distribution
    \item Kernel recombination for batch sampling
    \item GP hyperparameter optimisation
    \item Evidence estimation
\end{enumerate}
We iterate the above four procedures until the evidence variance reaches plateau. Only the first training procedure is different from the original BASQ \citep{adachi2022fast}.

The subsampling is to sample from the prior distribution to construct the empirical measure. As the kernel recombination is to select the sparse sample set from subsamples that can minimise the integral variance, subsamples should be sampled from prior but well overlapped from the higher predictive variance of GP $\ell(x)$. \cite{adachi2022fast} adopted uncertainty sampling for faster convergence, which samples from predictive variance $\sigma_\ell(x)$ and corrected to prior distribution via importance sampling, as such:
\begin{align*}
 g_\text{prop}(\Theta) &:= (1 - r) \mu_g(\Theta) + r \tilde A(\Theta), \quad 0 \leq r \leq 1 \\
 \text{w}_\text{IS}(\Theta) &:= \mu_g(\Theta) / g_\text{prop}(\Theta),\\
 \tilde A(\Theta) &:=  \sigma_g(\Theta) \pi^\prime(\Theta) / Z_{\tilde A},\\
 Z_{\tilde A} &:= \int_\Xi \sigma_g(\Theta) d\pi^\prime(\Theta),\\
 \sigma_g(\Theta) &:= \text{diag} \left[ \sigma_g(\Theta, \Theta) \right].
\end{align*}
We wish to adopt the same strategy for a four-layered GP, but the log-warp layer hinders the application. The predictive variance of the original BASQ can be analytically translated into the mixture of Gaussian with Gaussian kernel because the squared Gaussian distribution is still Gaussian. However, the exponentiated Gaussian is no more Gaussian, which becomes a log-normal distribution. As such, we cannot take the same strategy which exploits the Gaussianity. Hence, we employ the heuristical method. The predictive variance is expected to be larger at the midpoints between the observed data points. Thus, sampling from the midpoints with half lengthscale of GP is expected to be good proposal distribution of sampling the uncertainty region, as such:
\begin{align*}
 g_\text{heur}(\Theta) &:= \sum_{r, s}^{N_\text{heur}} w_{r,s}^\text{heur}\mathcal{N}\left(\Theta; \Theta^\text{mid}_{r,s}, \frac{\textbf{W}_\text{length}}{2} \right),\\
 \Theta^\text{mid}_{r,s} &:= \frac{\Theta_r + \Theta_s}{2},\\
 w_{r,s}^\text{heur} &:= \frac{\sigma_g( \Theta^\text{mid}_{r,s}) \pi^\prime (\Theta^\text{mid}_{r,s})}{\sum_{r,s}^{N_\text{heur}} \sigma_g(\Theta^\text{mid}_{r,s}) \pi^\prime (\Theta^\text{mid}_{r,s})},\\
\end{align*}
where $\Theta_r, \Theta_s \in \boldsymbol\Theta$ are the observed parameters, $\textbf{W}_\text{length}$ is the diagonal covariance matrix whose diagonal elements are the lengthscales of each dimension. Supersampling from this offers the uncertainty sampling, as such:
\begin{align*}
    \Theta_t^\text{super} &\sim  g_\text{heur}(\Theta) \in \mathbb{R}^{N_\text{super}},\\
    Z_{\tilde A} &= \int \sigma_g(\Theta) \frac{\pi^\prime(\Theta)}{g_\text{heur}(\Theta)}dg_\text{heur}(\Theta),\\
    &\approx \frac{1}{N_\text{super}} \sum_t^{N_\text{super}} \sigma_g(\Theta_t^\text{super}) \frac{\pi^\prime(\Theta_t^\text{super})}{g_\text{heur}(\Theta_t^\text{super})},\\
    w^\text{super} &:= \tilde A(\Theta_t^\text{super}) / g_\text{heur}(\Theta_t^\text{super}).
\end{align*}
Sequential Monte Carlo \citep{kitagawa1993monte} permits to sample from $\tilde A(\Theta)$.

\subsection{Ablation study of layered GPs}
\begin{table}
  \caption{Ablation study of warped layers}
  \label{tab:ablation}
  \centering
  \renewcommand{\arraystretch}{1.7}
    \begin{tabular}{ccc|cc}\toprule
    log & square-root & scaling & LEM & LEV\\\midrule
    \ding{52} & & & overflow & overflow\\
    & \ding{52} & & overflow & overflow\\
    & & \ding{52} & 361.8172 & -11.90735\\
    \ding{52} & & \ding{52} & 677.8633 & -21.86860\\
    & \ding{52} & \ding{52} & 449.6425 & -13.13063\\
    \ding{52} & \ding{52} & \ding{52} & \textbf{703.6569} & -27.31169\\\bottomrule
    \end{tabular}
\end{table}

We discuss the efficacy of four-layered GP by comparing the results of evidence inference for the easy case introduced in Table \ref{tab:easy}. We compared the following six configurations in Table \ref{tab:ablation}. The ground truth of LEM is estimated via exhaustive nested sampling with millions of samples until convergence, which yields 703.7285. The ablation study shows that the four-layered GPs can estimate the most accurate LEV of all compared configurations. GPs without the scaling layer reached the overflow limit, which returned a positive infinite value. GPs without the logarithmic layer scored the lower log evidence mean because the surrogate model cannot accommodate the wide dynamic range. Scaled GP with only log warp results was the second best. However, the non-exponentiated prior struggled to find the MAP location. As such, the four-layered GP, employing all features, was the performant.

\section{Identifiability derivation}\label{sec:ident}
\subsection{Hyperbolic secant distribution identities}
\begin{align}
\int_{-\infty}^\infty \text{sech}\left(x \right) dx
&= \pi, \label{eq:B1}\\
\int_{-\infty}^\infty \text{sech}\left( \frac{x - a}{b} \right) dx
&= \frac{\pi}{b}, \label{eq:B2}\\
\int_{-\infty}^\infty \text{sech}\left(x \right)
\ln \text{sech}\left(x \right)
dx
&= -\pi \ln2, \label{eq:B3}\\
\int_{-\infty}^\infty \text{sech}\left(x \right)
\text{sech}\left(x - a\right)
dx
&= 2a \text{csch}(a), \label{eq:B4}\\
\int_{-\infty}^\infty \text{sech}\left(x \right)^2 dx
&= 2. \label{eq:B5}
\end{align}

\subsection{SNR derivation}
\label{sec:snr}
\begin{align*}
    \text{SNR} &:= \ln\frac{\mathbb{V}\text{ar}_{P(\ln \omega)}[\text{Im}[Z]]}{\sigma^2_\text{noise}},\\
    \mathbb{V}\text{ar}_{P(\ln \omega)}[\text{Im}[Z]] &= \mathbb{E}_{P(\ln \omega)}[\text{Im}[Z]^2] - \mathbb{E}_{P(\ln \omega)}[\text{Im}[Z]]^2,\\
    \mathbb{E}_{P(\ln \omega)}[\text{Im}[Z]] &= \int_\Omega
    \text{Im}[Z](\ln\omega) \text{d}P(\ln \omega),\\
    &= \frac{\exp(r_\text{total}) \pi (1 - r_0)}{2(b-a)},\\
    \mathbb{E}_{P(\ln \omega)}[\text{Im}[Z]^2]
    &= \frac{\exp(2r_\text{total})(1 - r_0)^2}{2(b-a)} A,
\end{align*}
where
\begin{align*}
    P(\ln \omega) &:= \mathcal{U}(\ln \omega; a, b),\\
    a, b &:= \text{min}[\ln \omega], \text{max}[\ln \omega],\\
    A &:= \sum_i^N \lambda_i^2 + \sum_{i,j}^N 2\lambda_i \lambda_j \Delta \tau_{ij} \text{csch}(\Delta \tau_{ij}),\\
    \Delta \tau_{ij} &:= \sigma_\omega (\tau_i^\text{std} - \tau_j^\text{std}).
\end{align*}

Eq. (\ref{eq:B2}) yields the analytical solution of the first expectation:
\begin{align*}
\mathbb{E}_{P(\ln \omega)}[\text{Im}[Z]] &= \int_\Omega
P(\text{Im}[Z] | \omega) dP(\omega),\\
&= \frac{\exp(r_\text{total}) \pi (1 - r_0)}{2(b-a)} \sum_{i=1}^N \frac{\lambda_i}{\pi}\\
&\quad \int_{-\infty}^\infty \text{sech}\left(\omega + \Delta \tau_{ij} \right) d\omega,\\
&= \frac{\exp(r_\text{total}) \pi (1 - r_0)}{2(b-a)} \sum_{i=1}^N \lambda_i,\\
&= \frac{\exp(r_\text{total}) \pi (1 - r_0)}{2(b-a)}.
\end{align*}

Eqs. (\ref{eq:B4}) - (\ref{eq:B5}) yield the analytical solution of the second expectation:
\begin{align*}
\mathbb{E}_{P(\ln \omega)}[\text{Im}[Z]^2]
&= \int_\Omega
P(\text{Im}[Z] | \omega)^2 dP(\omega),\\
&= \frac{1}{b-a}\left[ \frac{\exp(r_\text{total}) \pi (1 - r_0)}{2} \right]^2\\
&\quad \int_{-\infty}^\infty
\left[
\sum_{i=1}^N \frac{\lambda_i}{\pi} \text{sech}\left(\omega + \Delta \tau_{ij} \right) 
\right]^2
d\omega,\\
&= \frac{1}{b-a}\left[ \frac{\exp(r_\text{total}) \pi (1 - r_0)}{2} \right]^2\\
&\quad \int_{-\infty}^\infty
\left[
\sum_{i}^N
\frac{\lambda_i^2}{\pi^2}
\text{sech}\left(\omega + \Delta \tau_{ij}\right)^2
+\right.\\
&\quad \left. \sum_{i,j}^N \frac{2 \lambda_i \lambda_j}{\pi^2}
\text{sech}\left(\omega \right)\text{sech}\left(\omega + \Delta \tau_{ij} \right) 
\right]
d\omega,\\
&= \frac{1}{b-a}\left[ \frac{\exp(r_\text{total}) \pi (1 - r_0)}{2} \right]^2\\
&\quad \left\{
\sum_i^N \frac{2\lambda_i^2}{\pi^2}
+
\sum_{i,j}^N
\frac{4\lambda_i \lambda_j}{\pi^2}
\Delta \tau_{ij}
\text{csch}\left(\Delta \tau_{ij} \right) 
\right\},\\
&= \frac{\exp(2r_\text{total})(1 - r_0)^2}{2(b-a)}\\
&\quad \left\{
\sum_i^N \lambda_i^2
+
\sum_{i,j}^N
2\lambda_i \lambda_j
\Delta \tau_{ij}
\text{csch}\left(\Delta \tau_{ij} \right) 
\right\}.
\end{align*}

\subsection{JS divergence derivation}
\label{sec:JS}
\subsubsection{Integral computation}
The JS divergence definition is as follows:
\begin{align*}
    \begin{split}
    \text{JS} &:= \frac{1}{2}\int_P \ln \left( \frac{P_i(x)}{M_{ij}(x)}
    \right) \text{d} P_i(x) \\
    &+ \frac{1}{2}\int_{P^\prime} \ln \left( \frac{P_j(x)}{M_{ij}(x)}
    \right) \text{d} P_j(x),\\
    \end{split}
\end{align*}
where
\begin{align*}
    M_{ij}(x) &:= \frac{1}{2}\Big(
    P_i(x)+P_j(x) \Big)
\end{align*}
To incorporate the information of weights, we adopt the following scaled hyperbolic secant distributions:
\begin{align*}
    P_i(\ln \omega) &:= \frac{\lambda_i}{\pi}\text{sech}\left[
    \lambda_i (\ln \omega + \sigma_\omega \tau_i^\text{std}) \right],\\
    P_j^\prime(\ln \omega) &:= \frac{\lambda_j}{\pi}\text{sech}\left[
    \lambda_j (\ln \omega + \sigma_\omega \tau_j^\text{std}) \right],
\end{align*}
where $\tau^\text{std}_j > \tau^\text{std}_i$. For efficient computation of the integrals, we can adopt the importance sampling. For simplicity, we show the calculation of the first term, given by:
\begin{align*}
\text{first term} &= \frac{1}{2}\int_P \frac{P_i(x)}{g_\text{JS}(x)} \ln \frac{P_i(x)}{M_{ij}(x)} \text{d} g_\text{JS}(x),\\
&\approx \frac{1}{2 N_\text{IS}} \sum_q^{N_\text{IS}} \frac{P_i(X^\text{IS}_q)}{g_\text{JS}(X^\text{IS}_q)} \ln \frac{P_i(X^\text{IS}_q)}{M_{ij}(X^\text{IS}_q)},\\
X^\text{IS}_q &\sim g_\text{JS}(x) \in \mathbb{R}^{N_\text{IS}},
\end{align*}
where
\begin{align*}
    g_\text{JS}(x) &:= \frac{1}{2N} \sum_i^N
    \frac{\lambda_i}{\pi}\text{sech}\left[ \lambda_i (x + \sigma_\omega \tau_i^\text{std}) \right]\\
    &\quad + \frac{1}{4\pi} \text{sech}\left[ 0.5(x + \sigma_\omega \lambda_i \tau_i^\text{std} + 0.5 \Delta_{ij}) \right],\\
    \Delta_{ij} &:= \sigma_\omega |\lambda_j \tau_j^\text{std} - \lambda_i \tau_i^\text{std}|,\\
\end{align*}
$g_\text{JS}(x)$ is a proposal distribution. As the logarithmic term is a subtraction of two hyperbolic secant distributions, the peak is estimated around the overlapped area, namely the midpoint of the two peaks $x + \sigma_\omega \lambda_i \tau_i^\text{std} + 0.5 \Delta_{ij}$. We can solve this integral via Monte Carlo integration. As sampling and evaluation of the probability density function of hyperbolic secant distribution are done within a millisecond order, computation with millions of samples for accuracy is not demanding.

\subsubsection{Noisy JS formulation}
The above computation assumes $P_i(\ln \omega)$ and $P_j(\ln \omega)$ probabilities are noise-free. In reality, the observed impedance is noisy, so we need to include the noise effect in the above formula to be more accurate. Note that the noise magnitude for impedance spectra is not $\sigma^2_\text{noise}$, but the exponentiated SNR.

We assume the noisy distribution as $P_i^\prime(\ln \omega)$, and the marginal probability can be obtained via marginalisation, as such:
\begin{align*}
    P_i^\prime(\ln \omega|\sigma^2_\text{n}) &\sim \mathcal{N}\Big(P^\prime_i; P_i(\ln \omega), \sigma^2_\text{n} \Big),\\
    P_i^\prime(\ln \omega) &= \int_Q P_i(\ln \omega|\sigma^2_\text{n}) \text{d} P_i(\sigma^2_\text{n}),
\end{align*}
where
\begin{align*}
    \sigma^2_\text{n} &= \exp(\text{SNR})\\
    P_i(\sigma^2_\text{n}) &= \text{LogNormal}(\sigma^2_\text{n}; \mu_\sigma, \sigma_\sigma).
\end{align*}
With regrad to the prior of $\sigma^2_\text{n}$, namely $P_i(\sigma^2_\text{n})$, we can adopt the same prior in Section \ref{sec:BayesInf}. That is, the prior for experimental noise is to extract the corresponding element in the prior $\pi(\Theta)$. So, the JS divergence with noise can be calculated by swapping both $P_i(x)$ and $P_j(x)$ with marginal $P^\prime_i(x)$ and $P^\prime_j(x)$.

\end{document}